\documentclass[notitlepage,a4paper,aps,prd,onecolumn,superscriptaddress,nofootinbib,groupedaddress,longbibliography]{revtex4-1}
\usepackage[utf8]{inputenc}
\usepackage[T1]{fontenc}
\usepackage{amsmath}
\usepackage{amsfonts}
\usepackage{amssymb}
\usepackage{mathtools}
\usepackage[colorlinks=true]{hyperref}
\usepackage[dvipsnames]{xcolor}
\usepackage{graphicx}
\usepackage[normalem]{ulem}

\newcommand{\order}[2]{\overset{\mathclap{\scriptscriptstyle #2}}{#1}\vphantom{#1}}

\newcommand{\lc}[1]{\overset{\circ}{#1}\vphantom{#1}}

\begin{document}

\title{Parameterized post-Newtonian limit of general teleparallel gravity theories}

\author{Ulbossyn Ualikhanova}
\email{ulbossyn.ualikhanova@ut.ee}
\affiliation{Laboratory of Theoretical Physics, Institute of Physics, University of Tartu, W. Ostwaldi 1, 50411 Tartu, Estonia}

\author{Manuel Hohmann}
\email{manuel.hohmann@ut.ee}
\affiliation{Laboratory of Theoretical Physics, Institute of Physics, University of Tartu, W. Ostwaldi 1, 50411 Tartu, Estonia}


\begin{abstract}
We derive the post-Newtonian limit of a general class of teleparallel gravity theories, whose action is given by a free function of three scalar quantities obtained from the torsion of the teleparallel connection. This class of theories is chosen to be sufficiently generic in order to include the $f(T)$ class of theories as well as new general relativity as subclasses. To derive its post-Newtonian limit, we first impose the Weitzenböck gauge, and then introduce a post-Newtonian approximation of the tetrad field around a Minkowski background solution. Our results show that the class of theories we consider is fully conservative, with only the parameters \(\beta\) and \(\gamma\) potentially deviating from their general relativity values. In particular, we find that the post-Newtonian limit of any $f(T)$ theory is identical to that of general relativity, so that these theories cannot be distinguished by measurements of the post-Newtonian parameters alone.
\end{abstract}

\maketitle

\section{Introduction}\label{sec:intro}
General relativity is challenged both by observations in cosmology and by its theoretical tensions with quantum theory. These challenges have led to the development of a plethora of modified gravity theories. While most of these theories take the most well-known formulation of general relativity in terms of the curvature of a Levi-Civita connection as their starting point, there exist other formulations which may serve as possible starting points for modifications~\cite{BeltranJimenez:2019tjy}. An important class of such modifications is based on the teleparallel equivalent of general relativity (TEGR)~\cite{Maluf:2013gaa}, and thus belongs to the class of teleparallel gravity theories~\cite{Moller:1961,Aldrovandi:2013wha,Maluf:2013gaa,Golovnev:2018red}. The characteristic feature of these theories is to employ a flat, metric-compatible connection, whose torsion mediates the gravitational interaction.

A large class of modified teleparallel gravity theories is obtained by assuming a gravitational Lagrangian of the form \(f(T)\)~\cite{Bengochea:2008gz,Linder:2010py}, where \(T\) is the torsion scalar appearing in the TEGR action~\cite{Maluf:2013gaa}. Various phenomenological and theoretical aspects of these theories have been investigated, including their cosmological dynamics~\cite{Cai:2015emx,Hohmann:2017jao,Nojiri:2017ncd} and perturbations~\cite{Golovnev:2018wbh}, gravitational waves~\cite{Bamba:2013ooa,Cai:2018rzd,Farrugia:2018gyz,Nunes:2018evm} and degrees of freedom from a Hamiltonian analysis~\cite{Li:2011rn,Ferraro:2018tpu,Ong:2018heg,Ferraro:2018axk,Bejarano:2019fii}. The rich phenomenology and generality of this class of gravity theories hence invite for further investigations of the class of a whole, studying further phenomenological aspects.

Another line of studies has been devoted to theories in which the three scalar quantities, which may be obtained from contractions of the torsion tensor, are treated separately. An early contender of this class is given by new general relativity~\cite{Hayashi:1979qx}, whose Lagrangian is simply the general linear combination of these three terms, and thus can be understood as derived from a general, local and linear constitutive relation~\cite{Itin:2016nxk,Itin:2018dru}. Several aspects of these theories have been studied, such as the equivalence principle~\cite{Shirafuji:1996im}, gravitational waves~\cite{Hohmann:2018jso} and Hamiltonian formulation~\cite{Blixt:2018znp}. Further relaxing the condition of linearity in the three scalar terms leads to an even more general class of teleparallel theories, whose action is given by a free function of three scalar quantities~\cite{Bahamonde:2017wwk,Hohmann:2018xnb}. This general class of teleparallel theories, which encompasses both the new relativity class of theories and the wide class of \(f(T)\) theories, will be the subject of our studies in this article.

While aiming to model the present observations in cosmology, any viable theory of gravity must of course also comply with observations on smaller scales, such as the solar system, orbiting pulsars and laboratory experiments. A commonly used framework which was developed for collectively deriving this local scale phenomenology is the parameterized post-Newtonian (PPN) formalism~\cite{Will:1993ns,Will:2014kxa,Will:2018bme}. It characterizes gravity theories by a set of ten parameters, which have been measured with high precision in various experiments. Because of its generality and the availability of numerous observations, the PPN formalism has become an important tool for assessing the viability of gravity theories.

In order to calculate the post-Newtonian limit of teleparallel theories of gravity, an adaptation of the classical PPN formalism to tetrad based theories is required. A possible adaptation can be derived from a similar approach to the post-Newtonian limit of scalar-tetrad theories~\cite{Hayward:1981bk}, by omitting the scalar field part. Further, it needs to be adapted to the covariant formulation of teleparallel gravity~\cite{Krssak:2015oua,Golovnev:2017dox,Krssak:2018ywd,Bejarano:2019fii}, which we will use in this article, and in which also a flat spin connection appears as a dynamical field. The purpose of this article is thus twofold. Our main aim is to put forward a general method for calculating the post-Newtonian limit of teleparallel gravity theories in their covariant formulation, by expanding the tetrad components in a pure spacetime basis and expressing them in terms of the post-Newtonian potentials and a number of constants, which are then determined by solving the field equations. The second aim is to use this general method in order to determine the post-Newtonian limit of a general class of teleparallel gravity theories~\cite{Bahamonde:2017wwk,Hohmann:2018xnb}. This class is chosen to be very generic, such as to encompass a large number of theories discussed in the literature, while at the same time being prototypical for applying our formalism to even more general theories.

Our work is in line with a number of previous studies of the post-Newtonian limit of the related class of Poincaré gauge theories. For a more restricted class of teleparallel theories, which is included in the class of theories we study here, it has been shown that post-Newtonian effects only occur at higher perturbation orders than the ones considered in the PPN formalism~\cite{Schweizer:1979up,Schweizer:1980vn,Smalley:1980em,Nitsch:1979qn}. More general classes of quadratic Poincaré gauge theories, in which both curvature and torsion are present, show deviations already at the PPN level, and may necessitate the use of additional PPN potentials and parameters beyond the standard formalism~\cite{Gladchenko:1990nw,Gladchenko:1994wu}. Note, however, that this will not be the case for the class of teleparallel gravity theories we consider in this article, for which the curvature of the considered connection vanishes identically.

The outline of this article is as follows. In section~\ref{sec:fieldvariables} we briefly review the dynamical variables and fields used in the covariant formulation of teleparallel gravity, and display the class of theories we consider, together with their action and field equations. In section~\ref{sec:pna} we review the basic ingredients of the post-Newtonian (PPN) formalism, and show how it can be adapted to the field variables relevant for teleparallel gravity. We employ this formalism in order to solve the field equations for a general post-Newtonian matter distribution in section~\ref{sec:ppl}. From this solution we obtain the post-Newtonian metric and PPN parameters in section~\ref{sec:pnparmet}, where we also compare our result with observations. Finally, in section~\ref{sec:examples} we discuss a number of specific examples. We end with a conclusion in section~\ref{sec:conclusion}.

In this article we use uppercase Latin letters \(A, B, \ldots = 0, \ldots, 3\) for Lorentz indices, lowercase Greek letters \(\mu, \nu, \ldots = 0, \ldots, 3\) for spacetime indices and lowercase Latin letters \(i, j, \ldots = 1, \ldots, 3\) for spatial indices. In our convention the Minkowski metric \(\eta_{AB}\) and \(\eta_{\mu\nu}\) has signature \((-,+,+,+)\).

\section{Field variables and their dynamics} \label{sec:fieldvariables}
We start with a brief review of the underlying geometry and dynamics of the theories we consider in this article. The fundamental variables in teleparallel theories of gravity, following their covariant formulation~\cite{Krssak:2015oua,Golovnev:2017dox,Krssak:2018ywd,Bejarano:2019fii}, are a tetrad $\theta^A{}_{\mu}$ and a curvature free Lorentz spin connection $\omega^A{}_{B\mu}$. We denote the inverse tetrad by \(e_A{}^{\mu}\), which satisfies \(\theta^A{}_{\mu}e_A{}^{\nu} = \delta_{\mu}^{\nu}\) and \(\theta^A{}_{\mu}e_B{}^{\mu} = \delta^A_B\). Via these variables one defines the metric
\begin{equation}\label{eqn:metric}
g_{\mu\nu} = \eta_{AB}\theta^A{}_{\mu}\theta^B{}_{\nu}
\end{equation}
and the torsion
\begin{equation}\label{eqn:torsionmunu}
T^{\rho}{}_{\mu\nu} = e_A{}^{\rho}\left(\partial_{\mu}\theta^A{}_{\nu} - \partial_{\nu}\theta^A{}_{\mu} + \omega^A{}_{B\mu}\theta^B{}_{\nu} - \omega^A{}_{B\nu}\theta^B{}_{\mu}\right)\,.
\end{equation}
To give dynamics to these fundamental field variables, we consider an action given by two parts,
\begin{equation} \label{eqn:totalaction}
S[\theta,\omega,\chi]=S_g[\theta,\omega]+S_m[\theta,\chi]\,,
\end{equation}
where \(S_g\) is the gravitational part, \(S_m\) is the matter part, and $\chi$ denotes an arbitrary set of matter fields. The variation of the matter action $S_m$ with respect to the tetrad $\theta^A{}_\mu$ can be written in the general form
\begin{equation}
\delta_\theta S_m = -\int_M \Theta_A{}^{\mu}\delta\theta^A{}_\mu\,\theta\,d^4x\,.
\end{equation}
Here $\theta$ is the determinant of the tetrad. Further, \(\Theta_A{}^{\mu}\) denotes the energy-momentum tensor, which we assume to be symmetric, $\Theta_{[\mu\nu]}=0$, by imposing local Lorentz invariance on the matter action. For the remainder of this article, we will treat the matter source as a perfect fluid, as discussed in detail in the section~\ref{sec:pna}. Also note that here we used the tetrad to change the index character, i.e., \(\Theta_{\mu\nu} = \theta^A{}_{\mu}g_{\nu\rho}\Theta_A{}^{\rho}\).

The gravitational part of the action $S_g$ is defined via the free function $\mathcal{F}$,
\begin{equation}\label{eqn:gravaction}
S_g[\theta,\omega] = \frac{1}{2\kappa^2}\int_M \mathcal{F}(\mathcal{T}_1,\mathcal{T}_2,\mathcal{T}_3)\,\theta\,d^4x,
\end{equation}
which depends on the three scalar quantities, which are parity-even and quadratic in the torsion, and take the form
\begin{equation}
\mathcal{T}_1 = T^{\mu\nu\rho}T_{\mu\nu\rho}\,, \quad
\mathcal{T}_2 = T^{\mu\nu\rho}T_{\rho\nu\mu}\,, \quad
\mathcal{T}_3 = T^\mu{}_{\mu\rho}T_\nu{}^{\nu\rho}\,.
\end{equation}
This action defines a generic class of teleparallel gravity theories, which has been discussed before in the literature~\cite{Bahamonde:2017wwk,Hohmann:2018xnb}, and shall serve both as a generic example and starting point for further extensions in future work.

By variation of the total action~\eqref{eqn:totalaction} with respect to the tetrad we find the field equations
\begin{multline}\label{eqn:generalfe}
\kappa^2\Theta_{\mu\nu} = \frac{1}{2}\mathcal{F}g_{\mu\nu} + 2\lc{\nabla}^\rho\left(\mathcal{F}_{,1}T_{\nu\mu\rho} + \mathcal{F}_{,2}T_{[\rho\mu]\nu} + \mathcal{F}_{,3}T^\sigma{}_{\sigma[\rho}g_{\mu]\nu}\right) + \mathcal{F}_{,1}T^{\rho\sigma}{}{}_{\mu}\left(T_{\nu\rho\sigma} - 2T_{[\rho\sigma]\nu}\right)\\
+ \frac{1}{2}\mathcal{F}_{,2}\left[T_{\mu}{}^{\rho\sigma}\left(2T_{\rho\sigma\nu} - T_{\nu\rho\sigma}\right) + T^{\rho\sigma}{}{}_{\mu}\left(2T_{[\rho\sigma]\nu} - T_{\nu\rho\sigma}\right)\right] - \frac{1}{2}\mathcal{F}_{,3}T^\sigma{}_{\sigma\rho}\left(T^\rho{}_{\mu\nu} + 2T_{(\mu\nu)}{}{}^{\rho}\right)\,,
\end{multline}
where $\mathcal{F}_{,i}=\partial\mathcal{F}/\partial\mathcal{T}_i$ with $i=1,2,3$ and \(\lc{\nabla}\) is the covariant derivative with respect to the Levi-Civita connection of the metric \(g_{\mu\nu}\). These are the field equations we will be solving in the remainder of this article. For this purpose we will make use of a post-Newtonian approximation of the teleparallel geometry, which will be detailed in the following section.

\section{Post-Newtonian approximation}\label{sec:pna}
The main tool we use in this article is the parameterized post-Newtonian (PPN) formalism ~\cite{Will:1993ns,Will:2014kxa,Will:2018bme}, which we briefly review in this section, taking into account that we intend to apply it to the class of extended teleparallel theories of gravity detailed in the preceding section. An important ingredient of the PPN formalism is the assumption that the matter which acts as the source of the gravitational field is given by a perfect fluid, whose velocity in a particular, fixed frame of reference is small, measured in units of the speed of light, and that all physical quantities relevant for the solution of the gravitational field equations can be expanded in orders of this velocity. In this section we discuss how this expansion in velocity orders proceeds for the quantities we need in our calculation in the following sections, in particular for the tetrad.

The starting point of our calculation is the energy-momentum tensor of a perfect fluid with rest energy density \(\rho\), specific internal energy \(\Pi\), pressure \(p\) and four-velocity \(u^{\mu}\), which is given by
\begin{equation}\label{eqn:tmunu}
\Theta^{\mu\nu} = (\rho + \rho\Pi + p)u^{\mu}u^{\nu} + pg^{\mu\nu}\,.
\end{equation}
The four-velocity \(u^{\mu}\) is normalized by the metric \(g_{\mu\nu}\), so that \(u^{\mu}u^{\nu}g_{\mu\nu} = -1\). We will now expand all dynamical quantities in orders \(\mathcal{O}(n) \propto |\vec{v}|^n\) of the velocity \(v^i = u^i/u^0\) of the source matter in a given frame of reference, starting with the field variables. We choose to work in the Weitzenböck gauge, and so we will set \(\omega^A{}_{B\mu} \equiv 0\). For the tetrad \(\theta^A{}_{\mu}\) we assume an expansion around a flat diagonal background tetrad \(\Delta^A{}_{\mu} = \mathrm{diag}(1, 1, 1, 1)\),
\begin{equation}\label{eqn:tetradexp}
\theta^A{}_{\mu} = \Delta^A{}_{\mu} + \tau^A{}_{\mu} = \Delta^A{}_{\mu} + \order{\tau}{1}^A{}_{\mu} + \order{\tau}{2}^A{}_{\mu} + \order{\tau}{3}^A{}_{\mu} + \order{\tau}{4}^A{}_{\mu} + \mathcal{O}(5)\,.
\end{equation}
Here we have used overscript numbers to denote velocity orders, i.e., each term \(\order{\tau}{n}^A{}_{\mu}\) is of order \(\mathcal{O}(n)\). Velocity orders beyond the fourth order are not considered and will not be relevant for our calculation.

For the tetrad perturbation \(\tau^A{}_{\mu}\) it will turn out to be more convenient to lower the Lorentz index using the Minkowski metric \(\eta_{AB}\) and convert it into a spacetime index using the background tetrad \(\Delta^A{}_{\mu}\), so that we introduce the perturbations
\begin{equation}
\tau_{\mu\nu} = \Delta^A{}_{\mu}\eta_{AB}\tau^B{}_{\nu}\,, \quad
\order{\tau}{n}_{\mu\nu} = \Delta^A{}_{\mu}\eta_{AB}\order{\tau}{n}^B{}_{\nu}\,.
\end{equation}
A detailed analysis shows that not all components of the tetrad field need to be expanded to the fourth velocity order, while others vanish due to Newtonian energy conservation or time reversal symmetry. The only relevant, non-vanishing components of the field variables we need to determine in this article are given by
\begin{equation}\label{eqn:ppnfields}
\order{\tau}{2}_{00}\,, \quad
\order{\tau}{2}_{ij}\,, \quad
\order{\tau}{3}_{0i}\,, \quad
\order{\tau}{3}_{i0}\,, \quad
\order{\tau}{4}_{00}\,.
\end{equation}
Using the expansion~\eqref{eqn:tetradexp} and the components listed above we can expand all geometric quantities appearing in the field equations up to their relevant velocity orders. This concerns in particular the metric, whose background solution follows from the diagonal background tetrad \(\Delta^A{}_{\mu}\) to be a flat Minkowski metric, \(\order{g}{0}_{\mu\nu} = \eta_{\mu\nu}\), and whose perturbation around this background is given by
\begin{equation}\label{eqn:metricexp}
\order{g}{2}_{00} = 2\order{\tau}{2}_{00}\,, \quad
\order{g}{2}_{ij} = 2\order{\tau}{2}_{(ij)}\,, \quad
\order{g}{3}_{0i} = 2\order{\tau}{3}_{(i0)}\,, \quad
\order{g}{4}_{00} = -(\order{\tau}{2}_{00})^2 + 2\order{\tau}{4}_{00}\,.
\end{equation}
For later use we also write out the relevant torsion components, which take the form
\begin{equation}\label{eqn:torsionexp}
\order{T}{2}^0{}_{0i} = \order{\tau}{2}_{00,i}\,, \quad
\order{T}{2}^i{}_{jk} = 2\delta^{il}\order{\tau}{2}_{l[k,j]}\,, \quad
\order{T}{3}^i{}_{0j} = \delta^{ik}(\order{\tau}{2}_{kj,0} - \order{\tau}{3}_{k0,j})\,, \quad
\order{T}{3}^0{}_{ij} = 2\order{\tau}{3}_{0[i,j]}\,, \quad
\order{T}{4}^0{}_{0i} = \order{\tau}{2}_{00}\order{\tau}{2}_{00,i} - \order{\tau}{3}_{0i,0} + \order{\tau}{4}_{00,i}\,,
\end{equation}
and which will be necessary for the decomposition of the field equations into velocity orders. Here we have made use of the additional assumption that the gravitational field is quasi-static, so that changes are only induced by the motion of the source matter. Time derivatives \(\partial_0\) of the tetrad components are therefore weighted with an additional velocity order \(\mathcal{O}(1)\).

Using the expansion~\eqref{eqn:metricexp} of the metric tensor we can now also expand the energy-momentum tensor~\eqref{eqn:tmunu} into velocity orders. For this purpose we must assign velocity orders also to the rest mass density, specific internal energy and pressure of the perfect fluid. Based on their orders of magnitude in the solar system one assigns velocity orders \(\mathcal{O}(2)\) to \(\rho\) and \(\Pi\) and \(\mathcal{O}(4)\) to \(p\). The energy-momentum tensor~\eqref{eqn:tmunu} can then be expanded in the form
\begin{subequations}\label{eqn:energymomentum}
\begin{align}
\Theta_{00} &= \rho\left(1 + \Pi + v^2 - 2\order{\tau}{2}_{00}\right) + \mathcal{O}(6)\,,\\
\Theta_{0j} &= -\rho v_j + \mathcal{O}(5)\,,\\
\Theta_{ij} &= \rho v_iv_j + p\delta_{ij} + \mathcal{O}(6)\,.
\end{align}
\end{subequations}
Finally, in order to expand also the gravitational side of the field equations~\eqref{eqn:generalfe}, we need to introduce a suitable expansion for the free function \(\mathcal{F}\) and its derivatives. For this purpose we use a Taylor expansion of the form
\begin{equation}
\mathcal{F}(\mathcal{T}_1, \mathcal{T}_2, \mathcal{T}_3) = \mathcal{F}(0, 0, 0) + \sum_{i = 1}^3\mathcal{F}_{,i}(0, 0, 0)\mathcal{T}_i + \mathcal{O}(\mathcal{T}^2)\,.
\end{equation}
Higher orders beyond the linear approximation will not be required. We further introduce the notation \(F = \mathcal{F}(0, 0, 0)\) and \(F_{,i} = \mathcal{F}_{,i}(0, 0, 0)\) for the constant Taylor coefficients. This well be used throughout the following sections.

\section{Expansion of the field equations and solution}\label{sec:ppl}
In order to discuss the post-Newtonian parameters, we need to expand the field equations to the required order in the perturbation and then make use of the post-Newtonian approximation. We will do so in the following sections. Further, we will make use of a generic ansatz for the tetrad perturbations, which consists of post-Newtonian potentials and constant coefficients, which we will also determine here by solving the field equations. We proceed order by order. The zeroth order, which corresponds to the background solution around which we expand, is discussed in section~\ref{ssec:ppn0}. We then solve for the second order in section~\ref{ssec:ppn2}, the third order in section~\ref{ssec:ppn3} and finally the fourth order in section~\ref{ssec:ppn4}.

\subsection{Background field equations}\label{ssec:ppn0}


We start our discussion with the zeroth order of the field equations~\eqref{eqn:generalfe}. From the expansion~\eqref{eqn:energymomentum} follows that at the zeroth velocity order the energy-momentum tensor vanishes, \(\order{\Theta}{0}_{\mu\nu} = 0\), so that we are left with solving the vacuum field equations. Inserting our assumed background values \(\order{\theta}{0}^A{}_{\mu} = \Delta^A{}_{\mu}\) for the tetrad into the respective field equations~\eqref{eqn:generalfe}, we find that they take the form
\begin{equation}
0 = \frac{1}{2}F\eta_{\mu\nu}\,.
\end{equation}
It thus follows that the field equations are solved at the zeroth order only for theories which satisfy \(F = 0\). This is a consequence of our assumption that the background solution is given by a flat Minkowski metric, which therefore excludes a cosmological constant. We will thus restrict ourselves to theories satisfying this restriction for the remainder of this article. This restriction will not be of importance for any actual phenomenology, since the effects of a non-vanishing cosmological constant in agreement with cosmological observations would be negligible on solar system scales.

\subsection{Second velocity order}\label{ssec:ppn2}
We continue with expanding the gravitational part $E_{\mu\nu}$ of the field equations~\eqref{eqn:generalfe} in the perturbation $\tau_{\mu\nu}$ at the second velocity order. The corresponding components take the form
\begin{subequations}\label{eqn:2order}
\begin{align}
\order{E}{2}_{00} &= -\left(2F_{,1} + F_{,2} + F_{,3}\right)\order{\tau}{2}_{00,ii} + 2F_{,3}\order{\tau}{2}_{i[i,j]j}\,,\label{eqn:e002}\\
\order{E}{2}_{ij} &= 4F_{,1}\order{\tau}{2}_{j[k,i]k}  +2 F_{,2}\left(\order{\tau}{2}_{i[k,j]k} + \order{\tau}{2}_{k[j,i]k}\right) + F_{,3}\left[2\order{\tau}{2}_{k[k,i]j} - \order{\tau}{2}_{00,ij}+\left(\order{\tau}{2}_{00,kk} + 2\order{\tau}{2}_{k[l,k]l}\right)\delta_{ij}\right]\,.\label{eqn:eij2}
\end{align}
\end{subequations}
It follows from their index structure that the tetrad components $\tau_{00}$, $\tau_{ij}$ should transform as a scalar and a tensor, respectively, under spatial rotations~\cite{Will:1993ns,Will:2018bme}. Further using their respective velocity orders and their relation to the source matter, we can write down an ansatz for the tetrad as
\begin{equation}\label{eqn:ansatz2}
\order{\tau}{2}_{00} = a_1U\,, \qquad
\order{\tau}{2}_{ij} = a_2U\delta_{ij} + a_3U_{ij}\,.
\end{equation}
Here $a_i$ (and also the later appearing $b_i$, $c_i$) are constant coefficients, which we will determine by solving the field equations and by imposing gauge conditions, while $U$ and $U_{ij}$ are post-Newtonian functionals of the matter variables. These functionals are related to the matter variables by the differential relations
\begin{equation}\label{eqn:functionals2}
\nabla^2 \chi = -2U\,, \qquad
U_{ij} = \chi_{,ij} + U\delta_{ij}\,, \qquad
\nabla^2 U = -4\pi\rho\,,
\end{equation}
where \(\nabla^2 = \delta^{ij}\partial_i\partial_j\) is the spatial Laplace operator of the flat background metric, and \(\chi\) is the so-called superpotential, which is auxiliary in the definition of \(U_{ij}\)~\cite{Will:1993ns}. For the sake of convenience, we will from now on rewrite the field equations making use of the shorthand notation $\order{\underline{E}}{n}_{\mu\nu}=\order{E}{n}_{\mu\nu}-\kappa^2\order{\Theta}{n}_{\mu\nu} = 0$. Then, inserting the appropriate ansatz~\eqref{eqn:ansatz2} for the tetrad into the field equations~\eqref{eqn:2order} at the second velocity order, and using the relations~\eqref{eqn:functionals2}, we obtain
\begin{subequations}\label{eqn:feans2}
\begin{align}
\order{\underline{E}}{2}_{00}&=-\left[\kappa^2-4\pi a_1(2F_{,1} + F_{,2} + F_{,3})+8\pi(a_2+a_3)F_{,3}\right]\rho\,,\label{eqn:feans002}\\
\order{\underline{E}}{2}_{ij}&=-\left[a_1F_{,3}-(a_2+a_3)(2F_{,1} + F_{,2} +2F_{,3})\right]\left(4\pi\delta_{ij}\rho+U_{,ij}\right)\,,\label{eqn:feansij2}
\end{align}
\end{subequations}
where we can see that the terms contained in square brackets in front of the post-Newtonian functionals must be zero, in order for the equations to be solved for arbitrary matter distributions. Further, note that we obtain only two independent equations, while our ansatz~\eqref{eqn:ansatz2} contains three free constants. This is a consequence of the gauge freedom, which is related to the diffeomorphism invariance of the theory. We thus may choose a gauge by supplementing the system with one additional equation. The standard PPN gauge mandates that the coefficient in front of \(U_{ij}\) vanishes, and so we make the gauge choice \(a_3 = 0\). Thus, we get for the coefficients
\begin{equation}\label{eqn:coeff2}
a_1= \frac{2F_{,1} + F_{,2} + 2F_{,3}}{(2F_{,1} + F_{,2})(2F_{,1} + F_{,2} + 3F_{,3})}\frac{\kappa^2}{4\pi},\qquad
a_2= \frac{F_{,3}}{(2F_{,1} + F_{,2})(2F_{,1} + F_{,2} + 3F_{,3})}\frac{\kappa^2}{4\pi},\qquad
a_3=0\,.
\end{equation}
We will subsequently use this second order solution in the remaining higher order field equations.

\subsection{Third velocity order}\label{ssec:ppn3}
At the third velocity order in the perturbation expansion we still work with linearized field equations, which are of the form
\begin{subequations}\label{eqn:3order}
\begin{align}
\order{E}{3}_{0i}  &= 2F_{,1}\left(\order{\tau}{2}_{ij,0j}-\order{\tau}{3}_{i0,jj}\right)+F_{,2}\left(\order{\tau}{2}_{ji,0j}-\order{\tau}{3}_{j0,ij}+2\order{\tau}{3}_{0[j,i]j}\right)+F_{,3}\left(\order{\tau}{2}_{jj,0i}-\order{\tau}{3}_{j0,ij}\right)\,,\label{eqn:e0i3}\\
\order{E}{3}_{i0}  &= 2F_{,1}\left(2\order{\tau}{3}_{0[j,i]j}-\order{\tau}{2}_{00,0i}\right)+F_{,2}\left(2\order{\tau}{3}_{[j|0|,i]j}+2\order{\tau}{2}_{[ij],0j}-\order{\tau}{2}_{00,0i}\right)+F_{,3}\left(2\order{\tau}{2}_{j[j,|0|i]}-\order{\tau}{2}_{00,0i}\right)\,. \label{eqn:ei03}
\end{align}
\end{subequations}
Observe that the components $\tau_{0i}$, $\tau_{i0}$ must behave as vectors under spatial rotations, which are of third velocity order, and so they can be expressed in terms of PPN potentials in the form
\begin{equation}\label{eqn:ansatz3}
\order{\tau}{3}_{i0} = b_1V_i+b_2W_i, \qquad \order{\tau}{3}_{0i} = b_3V_i+b_4W_i\,,
\end{equation}
with the PPN vector potentials satisfying
\begin{equation}
\nabla^2 V_{i} = -4\pi\rho v_i\,, \qquad
\nabla^2 W_{i} = -4\pi\rho v_i + 2 U_{,0i}\,.
\end{equation}
In this case, proceeding analogously to the equation~\eqref{eqn:feans2}, we obtain the third order field equations
\begin{subequations}\label{eqn:feans3}
\begin{align}
\order{\underline{E}}{3}_{i0}&= \left[\kappa^2+4\pi(b_1+b_2)F_{,2}+8\pi(b_3+b_4)F_{,1} \right]\left(\rho v_i-\frac{U_{,0i}}{4\pi}\right)\,,\label{eqn:feansi03}\\
\order{\underline{E}}{3}_{0i}&= \left[\kappa^2+8\pi(b_1+b_2)F_{,1}+4\pi(b_3+b_4)F_{,2}\right]\rho v_i  \\  \nonumber
&\phantom{=}+\left[(b_1-b_2)F_{,3}-b_2(4F_{,1}+F_{,2})+(b_1-b_3-b_4)F_{,2}+\frac{\kappa^2}{4\pi}\frac{F_{,3}}{2F_{,1} + F_{,2}}\right]U_{,0i}\,.\label{eqn:feans00i3}
\end{align}
\end{subequations}
We see that we obtain three independent equations, given by the vanishing of the square brackets, for the four coefficients \(b_1, \ldots, b_4\). This is again a consequence of the gauge invariance which we encountered also for the second order equations~\eqref{eqn:feans2} and coefficients~\eqref{eqn:ansatz2}. We could thus fix the gauge also here by adding one more equation. However, we will proceed differently in this case, and leave one of the constant coefficients undetermined at this stage. The reason for this will become clear at the fourth velocity order, where this free constant will allow us to choose the standard PPN gauge by eliminating one more PPN potential. Choosing \(b_4 = b_0\) as the undetermined parameter we find
\begin{equation}\label{eqn:coeff3}
b_1 = -\frac{1}{(2F_{,1} + F_{,2})}\frac{\kappa^2}{4\pi}\,, \qquad
b_2 = 0\,, \qquad
b_3 = -b_0-\frac{1}{(2F_{,1} + F_{,2})}\frac{\kappa^2}{4\pi}\,, \qquad
b_4 = b_0\,.
\end{equation}
Again, we will make use of this (now only partial) solution in the fourth order equations, which we address next.

\subsection{Fourth velocity order}\label{ssec:ppn4}
Finally, for the fourth order we find that we need to consider only certain components of the field equations and linear combinations thereof. In particular, we need the time component
\begin{equation}\label{eqn:4order}
\begin{split}
\order{E}{4}_{00} &= (2F_{,1} + F_{,2} + F_{,3})\left[-\order{\tau}{4}_{00,ii}+\order{\tau}{3}_{0i,0i}+\order{\tau}{2}_{00}\order{\tau}{2}_{00,ii}+2\order{\tau}{2}_{ij}\order{\tau}{2}_{00,ij}+\order{\tau}{2}_{00,i}\left(\order{\tau}{2}_{(ij),j}-\frac{\order{\tau}{2}_{00,i}}{2}-\order{\tau}{2}_{jj,i}\right)\right]\\
&+2F_{,1}\order{\tau}{2}_{ij,k}\order{\tau}{2}_{i[k,j]}-F_{,2}\order{\tau}{2}_{ij,k}\left(\order{\tau}{2}_{k[j,i]}+\order{\tau}{2}_{j[i,k]}\right)+\frac{F_{,3}}{2}\left(\order{\tau}{2}_{ij,i}\order{\tau}{2}_{kj,k}+\order{\tau}{2}_{ii,j}\order{\tau}{2}_{kk,j}+ 2\order{\tau}{2}_{ij,i}\order{\tau}{2}_{jk,k}\right)\\
&+2F_{,3}\left[\order{\tau}{4}_{i[i,j]j}+\order{\tau}{2}_{ij,k}\order{\tau}{2}_{j[k,i]}+2\order{\tau}{2}_{00}\order{\tau}{2}_{i[j,i]j}-\order{\tau}{2}_{ii,j}\order{\tau}{2}_{(jk),k}+\order{\tau}{2}_{ij}\left(\order{\tau}{2}_{j[k,i]k}+\order{\tau}{2}_{k(i,j)k}-\order{\tau}{2}_{kk,ij}\right)\right]
\end{split}
\end{equation}
and the trace of the spatial part of the field equations
\begin{equation}\label{eqn:4iiorder}
\begin{split}
\order{E}{4}_{ii}&=2(2F_{,1} + F_{,2} + 2F_{,3})\left(\order{\tau}{2}_{i[i,j]}\order{\tau}{2}_{jk,k}-\order{\tau}{4}_{i[i,j]j}\right)-2(F_{,1} + F_{,2} + F_{,3})\order{\tau}{2}_{ij,k}\order{\tau}{2}_{jk,i}-(2F_{,1} + F_{,2})\order{\tau}{2}_{ij}\order{\tau}{2}_{ij,kk}\\
&+2F_{,3}\left[\order{\tau}{4}_{00,ii}-\order{\tau}{3}_{0i,0i}-\order{\tau}{2}_{00,i}\order{\tau}{2}_{ij,j}+\order{\tau}{2}_{ii}\order{\tau}{2}_{jk,jk}-\order{\tau}{2}_{ij}\order{\tau}{2}_{jk,ik}+\order{\tau}{2}_{ji}\order{\tau}{2}_{ij,kk}-\order{\tau}{2}_{kk}\order{\tau}{2}_{ii,jj}+\order{\tau}{2}_{00,ii}\left(\order{\tau}{2}_{00}+\order{\tau}{2}_{jj}\right)\right]\\
&+(2F_{,1} + F_{,2} + 3F_{,3})\left[\order{\tau}{2}_{ii,00}-\order{\tau}{3}_{i0,i0}+2\order{\tau}{2}_{00,i}\order{\tau}{2}_{j[j,i]}+2\order{\tau}{2}_{ij}\left(\order{\tau}{2}_{kk,ij}-\order{\tau}{2}_{k(i,j)k}\right)\right]+\frac{1}{2}(2F_{,1} + F_{,2} + F_{,3})\order{\tau}{2}_{00,i}\order{\tau}{2}_{00,i}  \\
&+F_{,1}\left[2\order{\tau}{2}_{ik}\order{\tau}{2}_{ij,jk}+2\order{\tau}{2}_{kj,i}\order{\tau}{2}_{ki,j}+\order{\tau}{2}_{ij,k}\left(\order{\tau}{2}_{ij,k}-3\order{\tau}{2}_{ik,j}\right)\right]+\frac{F_{,2}}{2}\left(\order{\tau}{2}_{ij,k}\order{\tau}{2}_{kj,i}+2\order{\tau}{2}_{ij}\order{\tau}{2}_{ik,jk} \right)- 3F_{,3}\order{\tau}{2}_{(ij)}\order{\tau}{2}_{00,ij}\\
&+(2F_{,1} + F_{,2} + \frac{3}{2}c_3)\left[\order{\tau}{2}_{ii,j}\left(2\order{\tau}{2}_{kj,k}-\order{\tau}{2}_{kk,j}\right)-\order{\tau}{2}_{ij,i}\order{\tau}{2}_{kj,k}\right]+\left(2F_{,1} + \frac{3}{2}F_{,2} + 2F_{,3}\right)\order{\tau}{2}_{ij,k}\order{\tau}{2}_{ji,k}\,.
\end{split}
\end{equation}
In order to determine the post-Newtonian metric, we need to solve these equations for the tetrad component \(\order{\tau}{4}_{00}\). Note that this component should transform as a scalar under rotations, and thus we can consider an ansatz of the form
\begin{equation}\label{eqn:ansatz4}
\order{\tau}{4}_{00} = c_1\Phi_1+c_2\Phi_2+c_3\Phi_3+c_4\Phi_4 + c_5U^2\,,
\end{equation}
with the fourth order scalar potentials
\begin{equation}
\nabla^2 \Phi_1 = -4\pi\rho v^2\,, \qquad
\nabla^2 \Phi_2 = -4\pi\rho U\,, \qquad
\nabla^2 \Phi_3 = -4\pi\rho \Pi\,, \qquad
\nabla^2 \Phi_4 = -4\pi p\,.
\end{equation}
Finally, to eliminate the spatial component \(\order{\tau}{4}_{ij}\) of the tetrad, which appears in the field equations~\eqref{eqn:4order} and~\eqref{eqn:4iiorder}, but is not relevant for our calculation, we make use of the linear combination
\begin{equation}
\order{\underline{E}}{4} = (2F_{,1} + F_{,2} + 2F_{,3})\order{\underline{E}}{4}_{00} + F_{,3}\order{\underline{E}}{4}_{ii}
\end{equation}
and find
\begin{equation}\label{eqn:feans4}
\begin{split}
\order{\underline{E}}{4}&=(2F_{,1} + F_{,2})(2F_{,1} + F_{,2} + 3F_{,3})\left\{2b_0U_{,00}+4\pi[c_1\rho v^2+(c_2+2c_5)\rho U+c_3\rho\Pi+c_4 p]-2c_5U_{,i}U_{,i}\right\} \\
&\phantom{=}+\frac{\kappa^2}{4\pi}(2F_{,1} + F_{,2} + 2F_{,3})\left(U_{,00}+\frac{\kappa^2\rho U}{2F_{,1} + F_{,2} + 3F_{,3}}\right)-3F_{,3}\kappa^2p-(2F_{,1} + F_{,2} + 3F_{,3})\kappa^2\rho v^2\\
&\phantom{=}-\kappa^2(2F_{,1} + F_{,2} + 2F_{,3})\left(\rho\Pi+\frac{\kappa^2}{32\pi^2}\frac{U_{,i}U_{,i}}{2F_{,1} + F_{,2}}\right)\,.
\end{split}
\end{equation}
In order to obtain the solution in the standard PPN gauge, the coefficient in front of the term \(U_{,00}\) must vanish, since it does not correspond to any of the terms in the ansatz~\eqref{eqn:ansatz4}, and would introduce a term violating the standard PPN gauge. Together with the remaining, independent terms we then find the six independent equations
\begin{subequations}
\begin{align}
4\pi(2F_{,1} + F_{,2})(2F_{,1} + F_{,2} + 3F_{,3})c_4 - 3\kappa^3F_{,3} &= 0\,,\\
4\pi(2F_{,1} + F_{,2})(2F_{,1} + F_{,2} + 3F_{,3})c_3 - \kappa^2(2F_{,1} + F_{,2} + 2F_{,3}) &= 0\,,\\
4\pi(2F_{,1} + F_{,2})(2F_{,1} + F_{,2} + 3F_{,3})(c_2 + 2c_5) + \frac{\kappa^4}{4\pi}\frac{2F_{,1} + F_{,2} + 2F_{,3}}{2F_{,1} + F_{,2} + 3F_{,3}} &= 0\,,\\
2(2F_{,1} + F_{,2})(2F_{,1} + F_{,2} + 3F_{,3})b_0 + \frac{\kappa^2}{4\pi}(2F_{,1} + F_{,2} + 2F_{,3}) &= 0\,,\\
(2F_{,1} + F_{,2} + 3F_{,3})[4\pi(2F_{,1} + F_{,2})c_1 - \kappa^2] &= 0\,,\\
-2(2F_{,1} + F_{,2})(2F_{,1} + F_{,2} + 3F_{,3})c_5 - \frac{\kappa^4}{32\pi^2}\frac{2F_{,1} + F_{,2} + 2F_{,3}}{2F_{,1} + F_{,2}} &= 0\,.
\end{align}
\end{subequations}
Solving these equations for the remaining six undetermined constants then yields their values
\begin{gather}
b_0=-\frac{2F_{,1} + F_{,2} + 2F_{,3}}{(2F_{,1} + F_{,2})(2F_{,1} + F_{,2} + 3F_{,3})}\frac{\kappa^2}{8\pi}\,,\qquad
c_1=\frac{1}{(2F_{,1} + F_{,2})}\frac{\kappa^2}{4\pi}\,,\nonumber \\
c_2=-\frac{(2F_{,1} + F_{,2} -3F_{,3})(2F_{,1} + F_{,2} + 2F_{,3})}{(2F_{,1} + F_{,2})^2(2F_{,1} + F_{,2} + 3F_{,3})^2}\frac{\kappa^4}{32\pi^2}\,,\qquad
c_3=\frac{2F_{,1} + F_{,2} + 2F_{,3}}{(2F_{,1} + F_{,2})(2F_{,1} + F_{,2} + 3F_{,3})}\frac{\kappa^2}{4\pi}\,,\label{eqn:coeff4}\\
c_4=\frac{3F_{,3}}{(2F_{,1} + F_{,2})(2F_{,1} + F_{,2} + 3F_{,3})}\frac{\kappa^2}{4\pi}\,,\qquad
c_5=-\frac{2F_{,1} + F_{,2} + 2F_{,3}}{(2F_{,1} + F_{,2})^2(2F_{,1} + F_{,2} + 3F_{,3})}\frac{\kappa^4}{64\pi^2}\,.\nonumber
\end{gather}
With this result we have fully solved the general field equations~\eqref{eqn:generalfe} at all velocity orders which are required to determine the PPN metric and hence the PPN parameters. This will be done in the following section.

\section{PPN metric and parameters}\label{sec:pnparmet}
Using the solution obtained in the previous section, we can now finally determine the PPN metric and hence the PPN parameters of the general class of teleparallel gravity theories we consider in this article. We will do so in three steps. In section~\ref{ssec:tetradpn} we briefly recall the relevant tetrad components, and display their solutions after inserting the constant coefficients we determined into the respective ansatzes. From these components we derive the metric components in section~\ref{ssec:pnmetric}. Finally, in section~\ref{ssec:pnparameters}, we read off the PPN parameters. We compare this result to observations in section~\ref{ssec:obs}, in order to obtain bounds on the class of theories we consider.

\subsection{Post-Newtonian tetrad}\label{ssec:tetradpn}
We start by briefly recalling the tetrad components, and displaying their solutions, from section~\ref{sec:ppl}. From the ansatz~\eqref{eqn:ansatz2} together with the solutions~\eqref{eqn:coeff2} for the constant coefficients we find the second order components
\begin{equation}\label{eqn:tau2}
\order{\tau}{2}_{00} = \frac{2F_{,1} + F_{,2} + 2F_{,3}}{(2F_{,1} + F_{,2})(2F_{,1} + F_{,2} + 3F_{,3})}\frac{\kappa^2}{4\pi}U\,, \qquad
\order{\tau}{2}_{ij} = \frac{F_{,3}}{(2F_{,1} + F_{,2})(2F_{,1} + F_{,2} + 3F_{,3})}\frac{\kappa^2}{4\pi}U\delta_{ij}\,.
\end{equation}
We then come to the third order ansatz~\eqref{eqn:ansatz3}, together with the solution~\eqref{eqn:coeff3} and the missing coefficient \(b_0\) in the solution~\eqref{eqn:coeff4}. This yields the components
\begin{equation}
\order{\tau}{3}_{i0} = -\frac{1}{2F_{,1} + F_{,2}}\frac{\kappa^2}{4\pi}V_i\,, \qquad
\order{\tau}{3}_{0i} = -\frac{\kappa^2}{8\pi}\frac{(2F_{,1} + F_{,2} + 4F_{,3})V_i + (2F_{,1} + F_{,2} + 2F_{,3})W_i}{(2F_{,1} + F_{,2})(2F_{,1} + F_{,2} + 3F_{,3})}\,.\label{eqn:tau3}
\end{equation}
Finally, we recall the ansatz~\eqref{eqn:ansatz4} for the only fourth order component we have to determine. With the solution~\eqref{eqn:coeff4} we find
\begin{multline}
\order{\tau}{4}_{00} = \frac{1}{2F_{,1} + F_{,2}}\frac{\kappa^2}{4\pi}\Phi_1- \frac{(2F_{,1} + F_{,2} - 3F_{,3})(2F_{,1} + F_{,2} + 2F_{,3})}{(2F_{,1} + F_{,2})^2(2F_{,1} + F_{,2} + 3F_{,3})^2}\frac{\kappa^4}{32\pi^2}\Phi_2+ \frac{2F_{,1} + F_{,2} + 2F_{,3}}{(2F_{,1} + F_{,2})(2F_{,1} + F_{,2} + 3F_{,3})}\frac{\kappa^2}{4\pi}\Phi_3\\
+\frac{3F_{,3}}{(2F_{,1} + F_{,2})(2F_{,1} + F_{,2} + 3F_{,3})}\frac{\kappa^2}{4\pi}\Phi_4 - \frac{2F_{,1} + F_{,2} + 2F_{,3}}{(2F_{,1} + F_{,2})^2(2F_{,1} + F_{,2} + 3F_{,3})}\frac{\kappa^4}{64\pi^2}U^2\,.\label{eqn:tau4}
\end{multline}
These are all tetrad components which are relevant to construct the post-Newtonian metric.

\subsection{Post-Newtonian metric}\label{ssec:pnmetric}
In the next step we calculate the post-Newtonian metric. For this purpose we insert the tetrad components displayed in section~\ref{ssec:tetradpn} into the metric expansion~\eqref{eqn:metricexp}. We start with the second order metric component
\begin{equation}
\order{g}{2}_{00} = \frac{2F_{,1} + F_{,2} + 2F_{,3}}{(2F_{,1} + F_{,2})(2F_{,1} + F_{,2} + 3F_{,3})}\frac{\kappa^2}{2\pi}U = 2GU\,,
\end{equation}
which follows immediately from the second order tetrad perturbation~\eqref{eqn:tau2}. Here we introduced the Newtonian gravitational constant \(G\). Solving the normalization condition \(G = 1\), as this is the conventional PPN choice of units, yields the relation
\begin{equation}
\kappa^2 = 4\pi\frac{(2F_{,1} + F_{,2})(2F_{,1} + F_{,2} + 3F_{,3})}{2F_{,1} + F_{,2} + 2F_{,3}}\,.
\end{equation}
Using this normalization we find for the remaining components
\begin{equation}
\order{g}{2}_{ij} = \frac{2F_{,3}}{2F_{,1} + F_{,2} + 2F_{,3}}U\delta_{ij}
\end{equation}
at the second order,
\begin{equation}
\order{g}{3}_{0i} = -\frac{6F_{,1} + 3F_{,2} + 10F_{,3}}{2(2F_{,1} + F_{,2} + 2F_{,3})}V_i - \frac{1}{2}W_i
\end{equation}
at the third order and finally
\begin{multline}
\order{g}{4}_{00} = \frac{1}{2F_{,1} + F_{,2} + 2F_{,3}}\bigg[-\frac{6F_{,1} + 3F_{,2} + 7F_{,3}}{2}U^2 + 2(2F_{,1} + F_{,2} + 3F_{,3})\Phi_1\\
-(2F_{,1} + F_{,2} - 3F_{,3})\Phi_2 + 2(2F_{,1} + F_{,2} + 2F_{,3})\Phi_3 + 6F_{,3}\Phi_4\bigg]
\end{multline}
at the fourth order. Further components will not be necessary in order to obtain the PPN parameters.

\subsection{Post-Newtonian parameters}\label{ssec:pnparameters}
By comparing the metric components shown in section~\ref{ssec:pnmetric} with the standard PPN form of the metric~\cite{Will:1993ns,Will:2018bme}, we find the PPN parameters for the theory as
\begin{equation}
\xi = \alpha_1 = \alpha_2 = \alpha_3 = \zeta_1 = \zeta_2 = \zeta_3 =\zeta_4 = 0\,,
\end{equation}
from which we deduce that there is no violation of the conservation of total energy-momentum, as well as no preferred frame or preferred location effects; theories of this type are called fully conservative. The only non-trivial result is given by the PPN parameters
\begin{equation}\label{eqn:bg}
\beta = \frac{6F_{,1} + 3F_{,2} + 7F_{,3}}{4(2F_{,1} + F_{,2} + 2F_{,3})}\,, \qquad
\gamma = \frac{F_{,3}}{2F_{,1} + F_{,2} + 2F_{,3}}\,.
\end{equation}
More expressively, we find that their deviation from the general relativity values \(\beta_{\text{GR}} = \gamma_{\text{GR}} = 1\) can be written in terms of a single constant \(\epsilon\) by defining
\begin{equation}\label{eqn:bgdev}
\beta - 1 = -\frac{\epsilon}{2}\,, \qquad
\gamma - 1 = -2\epsilon\,, \qquad
\epsilon = \frac{2F_{,1} + F_{,2} + F_{,3}}{2(2F_{,1} + F_{,2} + 2F_{,3})}\,.
\end{equation}
In particular, we obtain \(\beta = \gamma = 1\) for \(2F_{,1} + F_{,2} + F_{,3} = 0\), so that theories satisfying these conditions are indistinguishable from general relativity by measurements of the PPN parameters. We will discuss this particular case later in section~\ref{sec:examples}, when we discuss specific examples.

\subsection{Comparison to observations}\label{ssec:obs}
For the discussion of experimental bounds it is important to take into account that the deviations~\eqref{eqn:bgdev} of the PPN parameters from their general relativity values are not independent. This fact is relevant for most measurements of the PPN parameters, where the result depends on a linear combination of the parameters, such as the perihelion shift of Mercury or the Nordtvedt effect~\cite{Will:2014kxa}. The latter is in particular remarkable, since from the values~\eqref{eqn:bg} follows \(4\beta - \gamma = 3\), so that the Nordtvedt parameter~\cite{Nordtvedt:1968qr,Nordtvedt:1968qs}
\begin{equation}
\eta_N = 4\beta - \gamma - 3 - \frac{10}{3}\xi - \alpha_1 + \frac{2}{3}\alpha_2 - \frac{2}{3}\zeta_1 - \frac{1}{3}\zeta_2
\end{equation}
vanishes identically, indicating the absence of the Nordtvedt effect independently of the theory under consideration. Hence, lunar laser ranging experiments searching for the Nordtvedt effect will not be affected, and are thus insensitive to the modifications we discuss here.

For measurements of the PPN parameter \(\gamma\) alone, the most stringent bound is obtained from the Cassini tracking experiment~\cite{Bertotti:2003rm}, which yields the bound
\begin{equation}
\gamma - 1 = -2\epsilon \leq (2.1 \pm 2.3) \cdot 10^{-5}\,.
\end{equation}
Comparable bounds on \(\epsilon\) may be obtained from solar system ephemeris, which yields bounds on both \(\gamma\) and \(\beta\)~\cite{Verma:2013ata}.

This concludes our discussion of the PPN parameters for a general teleparallel theory. To illustrate our results, we will present the most commonly encountered examples in the following section.

\section{Examples}\label{sec:examples}
We now apply the general result we derived in the previous sections to a number of example theories. We start with a simple rewriting of the gravitational Lagrangian in its axial, vector and tensor parts in section~\ref{ssec:axvecten}. In section~\ref{ssec:ngr}, we then consider new general relativity, in which the general function \(\mathcal{F}\) is replaced by a linear function of its three arguments. In section~\ref{ssec:fT} we finally consider the \(f(T)\) class of theories, where \(f\) is a function depending on the TEGR torsion scalar only.

\subsection{$\mathcal{G}(T_{\text{ax}}, T_{\text{vec}}, T_{\text{ten}})$ theories}\label{ssec:axvecten}
We begin by noting that the theory of gravity given in~\cite{Bahamonde:2017wwk} is identical to the class of theories we discussed here, since its action is of the same form
\begin{equation}
\mathcal{F}(\mathcal{T}_1,\mathcal{T}_2,\mathcal{T}_3) = \mathcal{G}(T_{\text{ax}}, T_{\text{vec}}, T_{\text{ten}})
\end{equation}
with the torsion components
\begin{equation}
T_{\text{ax}}=\frac{1}{18}(\mathcal{T}_1-2\mathcal{T}_2)\,, \quad
T_{\text{ten}}=\frac{1}{2}(\mathcal{T}_1+\mathcal{T}_2-\mathcal{T}_3)\,, \quad
T_{\text{vec}}=\mathcal{T}_3\,,
\end{equation}
which are fully equivalent for expressing the action. It follows that the Taylor coefficients
\begin{equation}
G = \mathcal{G}|_{T = 0}\,, \quad
G_{,a} = \left.\frac{\partial\mathcal{G}}{\partial T_{\text{ax}}}\right|_{T = 0}\,, \quad
G_{,t} = \left.\frac{\partial\mathcal{G}}{\partial T_{\text{ten}}}\right|_{T = 0}\,, \quad
G_{,v} = \left.\frac{\partial\mathcal{G}}{\partial T_{\text{vec}}}\right|_{T = 0}\,, \quad
\end{equation} are related by
\begin{equation}
F = G\,, \quad
F_{,1} = \frac{1}{18}G_{,a} + \frac{1}{2}G_{,t}\,, \quad
F_{,2} = -\frac{1}{9}G_{,a} + \frac{1}{2}G_{,t}\,, \quad
F_{,3} = G_{,v} - \frac{1}{2}G_{,t}\,.
\end{equation}
Note in particular that \(G_{,a}\) drops out whenever \(F_{,1}\) and \(F_{,2}\) appear only in the combination \(2F_{,1} + F_{,2}\). Hence, the axial part does not contribute to the deviation~\eqref{eqn:bgdev} of the PPN parameters from their general relativity values, since
\begin{equation}
\epsilon = \frac{G_{,v} + G_{,t}}{4G_{,v} + G_{,t}}
\end{equation}
contains only vectorial and tensorial parts. This agrees with earlier findings, that purely axial modifications show up only in higher post-Newtonian orders than considered in the PPN formalism~\cite{Schweizer:1979up,Schweizer:1980vn,Smalley:1980em,Nitsch:1979qn}.

\subsection{New general relativity}\label{ssec:ngr}
Next, we consider the new general relativity (NGR) class of teleparallel gravity theories~\cite{Hayashi:1979qx}. Its Lagrangian is given by the general linear combination
\begin{equation}
\mathcal{F}(\mathcal{T}_1,\mathcal{T}_2,\mathcal{T}_3) = t_1\mathcal{T}_1 + t_2\mathcal{T}_2 + t_3\mathcal{T}_3
\end{equation}
with constant coefficients \(t_i\). It thus follows immediately that the Taylor coefficients are given by $F=0$ and $F_{,i} = t_i$, $i=1,2,3$. The deviation~\eqref{eqn:bgdev} of the PPN parameters is thus given by
\begin{equation}
\epsilon = \frac{2t_1 + t_2 +t_3}{2(2t_1 + t_2 + 2t_3)}\,.
\end{equation}
This result agrees with the values obtained for \(\beta\) and \(\gamma\) in the original presentation~\cite{Hayashi:1979qx} of the theory.

\subsection{$f(T)$ theories}\label{ssec:fT}
Another important class of theories which is covered by the calculations we present in this article is given by the so-called \(f(T)\) class of theories, whose Lagrangian is given by
\begin{equation}
\mathcal{F}(\mathcal{T}_1,\mathcal{T}_2,\mathcal{T}_3) = f(T)\,, \quad
T = \frac{1}{4}\mathcal{T}_1 + \frac{1}{2}\mathcal{T}_2 - \mathcal{T}_3\,.
\end{equation}
Here \(T\) is the torsion scalar which constitutes the Lagrangian of the teleparallel equivalent of general relativity (TEGR)~\cite{Maluf:2013gaa}. For the Taylor coefficients we find \(F = f(0)\), so that at the zeroth order we get the condition $F=f(0)=0$. The remaining Taylor coefficients are given by $F_{,1} = \frac{1}{4}f'(0)$, $F_{,2} = \frac{1}{2}f'(0)$ and $F_{,3} = -f'(0)$. As a consequence, we find that the deviation~\eqref{eqn:bgdev} of the PPN parameters from their general relativity values vanishes identically, \(\epsilon = 0\), for any theories of this class. Hence, we find that any \(f(T)\) type theories cannot be distinguished from general relativity by their PPN parameters.

\section{Conclusion}\label{sec:conclusion}
We derived the post-Newtonian limit of a general class of teleparallel gravity theories, whose action is given by a Lagrange function depending on three scalar quantities formed from the parity-even contractions of the torsion tensor~\cite{Bahamonde:2017wwk,Hohmann:2018xnb}. We found that the post-Newtonian limit of these theories is fully determined by a single constant, which is calculated from four Taylor coefficients of the Lagrange function at the zeroth and first order. The zeroth order, which plays the role of a cosmological constant, must be set to zero to achieve consistency between the background (vacuum) field equations and the post-Newtonian ansatz of a flat Minkowski background (or at least sufficiently small such as not to affect the solar system dynamics). The post-Newtonian parameters are then fully determined by the first order Taylor coefficients. We displayed these coefficients in two different representations, both through the canonical contractions of the torsion tensor and its axial-vector-tensor decomposition.

Our results show that the class of theories we considered is fully conservative in the sense that it does not exhibit any preferred frame or preferred location effects, or violation of energy-momentum conservation, which is reflected by the fact that only the PPN parameters \(\gamma\) and \(\beta\) potentially deviate from their general relativity values. Further, due to the aforementioned fact that deviations of the PPN parameters from their general relativity values are governed by a single combination of the constant Taylor coefficients, large parts of the parameter space of possible theories are left with a post-Newtonian limit which is identical to that of general relativity, so that these theories are indistinguishable by solar system experiments at the respective post-Newtonian order. Further, we found that the Nordtvedt effect is absent in the whole class of theories we considered.

We then applied our findings to two particular subclasses of theories: new general relativity~\cite{Hayashi:1979qx} and $f(T)$ gravity~\cite{Bengochea:2008gz,Linder:2010py}. In the former case the aforementioned Taylor coefficients are given by the three constant parameters which determine the new general relativity action, and our findings agree with the original calculation of \(\gamma\) and \(\beta\) from a static, spherically symmetric ansatz~\cite{Hayashi:1979qx}. In the latter case we find that the post-Newtonian parameters are identical to those of general relativity, so that any $f(T)$ gravity theory is consistent with solar system observations.

Our work invites for numerous generalizations and extensions. In particular, one may consider more general theories, for example derived from a general constitutive relation~\cite{Hohmann:2017duq}, possibly including also parity-odd terms. Another possibility is to include a coupling to scalar fields~\cite{Hohmann:2018rwf,Hohmann:2018vle,Hohmann:2018dqh,Hohmann:2018ijr,Emtsova:2019qsl,Flathmann:2019khc}, up to Horndeski-like teleparallel theories~\cite{Bahamonde:2019shr,Hohmann:2019gmt}. This would extend previous calculations of the PPN parameters for specific theories in this class~\cite{Li:2013oef,Chen:2014qsa,Sadjadi:2016kwj}. Further, taking inspiration from the so-called trinity of gravity~\cite{BeltranJimenez:2019tjy}, one may consider extensions to the symmetric teleparallel equivalent of gravity~\cite{Nester:1998mp}, and apply the parameterized post-Newtonian formalism to generalized theories based on the symmetric teleparallel geometry~\cite{BeltranJimenez:2017tkd,Conroy:2017yln,BeltranJimenez:2018vdo,Jarv:2018bgs,Runkla:2018xrv}. Another possible extension would be studying the motion of compact objects at higher orders in the post-Newtonian expansion, in order to derive the emitted gravitational waves~\cite{Blanchet:2013haa}.

\begin{acknowledgments}
UU gratefully acknowledges mobility funding from the European Union Framework Programme Horizon 2020 through COST Action: CA15117, and thanks Mariafelicia De Laurentis for hospitality and supervision during a research visit in Naples. MH and UU gratefully acknowledge the full financial support by the Estonian Research Council through the Personal Research Funding project PRG356 and by the European Regional Development Fund through the Center of Excellence TK133 ``The Dark Side of the Universe''.
\end{acknowledgments}

\bibliography{../teleppn}
\end{document}